\documentclass[aps,prl,twocolumn,showpacs,showkeys,tightenlines,superscriptaddress]{revtex4} 
\usepackage{epsfig,rotating}

\newcommand{\nuc}[2]{\hbox{$^{#1}$#2}}

\begin{document}

\title{Measurement of excited states in $^{40}$Si and evidence for weakening of the $N=28$ shell gap}

\author{C.\,M.~Campbell}
   \affiliation{National Superconducting Cyclotron Laboratory,
     Michigan State University,
     East Lansing, Michigan 48824}
   \affiliation{Department of Physics and Astronomy,
     Michigan State University,
     East Lansing, Michigan 48824}
\author{N.~Aoi}
   \affiliation{The Institute of Physical and Chemical Research,
     2-1 Hirosawa, Wako, Saitama 351-0198, Japan}
\author{D.~Bazin}
   \affiliation{National Superconducting Cyclotron Laboratory,
     Michigan State University,
     East Lansing, Michigan 48824}
\author{M.\,D.~Bowen}
   \affiliation{National Superconducting Cyclotron Laboratory,
     Michigan State University,
     East Lansing, Michigan 48824}
   \affiliation{Department of Physics and Astronomy,
     Michigan State University,
     East Lansing, Michigan 48824}
\author{B.\,A.~Brown}
   \affiliation{National Superconducting Cyclotron Laboratory,
     Michigan State University,
     East Lansing, Michigan 48824}
   \affiliation{Department of Physics and Astronomy,
     Michigan State University,
     East Lansing, Michigan 48824}
\author{J.\,M.~Cook}
   \affiliation{National Superconducting Cyclotron Laboratory,
     Michigan State University,
     East Lansing, Michigan 48824}
   \affiliation{Department of Physics and Astronomy,
     Michigan State University,
     East Lansing, Michigan 48824}
\author{D.\,C.~Dinca}
   \affiliation{National Superconducting Cyclotron Laboratory,
     Michigan State University,
     East Lansing, Michigan 48824}
   \affiliation{Department of Physics and Astronomy,
     Michigan State University,
     East Lansing, Michigan 48824}
\author{A.~Gade}
   \affiliation{National Superconducting Cyclotron Laboratory,
     Michigan State University,
     East Lansing, Michigan 48824}
   \affiliation{Department of Physics and Astronomy,
     Michigan State University,
     East Lansing, Michigan 48824}
\author{T.~Glasmacher}
   \affiliation{National Superconducting Cyclotron Laboratory,
     Michigan State University,
     East Lansing, Michigan 48824}
   \affiliation{Department of Physics and Astronomy,
     Michigan State University,
     East Lansing, Michigan 48824}
\author{M.~Horoi}
   \affiliation{Physics Department,
     Central Michigan University, 
     Mount Pleasant, Michigan 48859}
\author{S.~Kanno}
   \affiliation{Department of Physics,
     Rikkyo University,
     3-34-1 Nishi-Ikebukuro, Toshima, Tokyo 171-8501, Japan}
\author{T.~Motobayashi}
   \affiliation{The Institute of Physical and Chemical Research,
     2-1 Hirosawa, Wako, Saitama 351-0198, Japan}
\author{W.\,F.~Mueller}
   \affiliation{National Superconducting Cyclotron Laboratory,
     Michigan State University,
     East Lansing, Michigan 48824}
\author{H.~Sakurai}
   \affiliation{The Institute of Physical and Chemical Research,
     2-1 Hirosawa, Wako, Saitama 351-0198, Japan}
\author{K.~Starosta}
   \affiliation{National Superconducting Cyclotron Laboratory,
     Michigan State University,
     East Lansing, Michigan 48824}
   \affiliation{Department of Physics and Astronomy,
     Michigan State University,
     East Lansing, Michigan 48824}
\author{H.~Suzuki}
   \affiliation{Department of Physics,
     University of Tokyo,
     Tokyo 1130033, Japan}
\author{S.~Takeuchi}
   \affiliation{The Institute of Physical and Chemical Research,
     2-1 Hirosawa, Wako, Saitama 351-0198, Japan}
\author{J.\,R.~Terry}
   \affiliation{National Superconducting Cyclotron Laboratory,
     Michigan State University,
     East Lansing, Michigan 48824}
   \affiliation{Department of Physics and Astronomy,
     Michigan State University,
     East Lansing, Michigan 48824}
\author{K.~Yoneda}
   \affiliation{The Institute of Physical and Chemical Research,
     2-1 Hirosawa, Wako, Saitama 351-0198, Japan}
\author{H.~Zwahlen}
   \affiliation{National Superconducting Cyclotron Laboratory,
     Michigan State University,
     East Lansing, Michigan 48824}
   \affiliation{Department of Physics and Astronomy,
     Michigan State University,
     East Lansing, Michigan 48824}

\date{\today}

\begin{abstract}
Excited states in \nuc{40}{Si} have been established by 
detecting $\gamma$-rays coincident with inelastic scattering and 
nucleon removal reactions on a liquid hydrogen target.  The low 
excitation energy, 986(5) keV, of the $2^+_1$ state provides 
evidence of a weakening in the $N=28$ shell closure in a neutron-rich 
nucleus devoid of deformation-driving proton collectivity.
\end{abstract}



\maketitle

The shell structure of the atomic nucleus provides one of the most 
important building blocks for the understanding of this correlated, 
fermionic many-body system \cite{cau05,bro01}. Nuclei with a closed shell 
of protons or neutrons are particularly stable \cite{mgm49,hjs49}. 
In neutron-rich exotic species, the shell structure, well-established 
close to stability, is modified; new magic numbers appear and some 
traditional shell gaps vanish \cite{war04,taka01}. Those changes in a regime 
of a pronounced imbalance between proton and neutron numbers are driven, 
for example, by the tensor force \cite{taka05} and the proton-neutron monopole 
interaction \cite{taka01,fed79,heyde85}.

A particularly fertile ground to study modifications to shell 
structure far from stability are neutron-rich nuclei with protons in the 
$sd$ shell and neutrons in the $fp$ shell ($\pi(sd)\nu(fp)$). The collapse 
of the $N=20$ shell closure in Ne, Na and Mg isotopes 
is actively studied three decades after its discovery
\cite{thib75,camp75,moto95,boris99,church05,neid05,trip05,ney05,iwa05}.  
Prior experimental 
studies and theoretical predictions suggest a weakening of the $N=28$ 
magic number \cite{sor93,gla97,wer9496,lala99}. Coulomb excitation of the 
$N=28$ nucleus \nuc{44}{S} implied enhanced collectivity \cite{gla97} and 
$\beta$-decay experiments have been used to infer a deformed ground state 
\cite{sor93}. The question of whether these observations are due to a breakdown 
of the $N=28$ magic number or the collapse of the $Z=16$ proton sub-shell gap 
at neutron number $N=28$ are much discussed in the literature 
\cite{reta97,cot98}.  A near-degeneracy of the $\pi1s_{1/2}$  and $\pi0d_{3/2}$ 
orbitals has been established experimentally for stable \cite{doll76} and 
neutron-rich nuclei \cite{fri05,cot06} at $N=28$.  However, observation of 
collectivity in these nuclei cannot be used to draw firm conclusions on 
the breakdown of $N=28$ because proton degeneracy also enhances 
collectivity approaching $N=28$.

To isolate changes in the $N=28$ shell closure, we studied excited states 
in \nuc{40}{Si}($Z=14$, $N=26$) where recent experiments indicate the $Z=14$ 
proton sub-shell gap remains large at $N=28$ \cite{fri05,cot06}.  
Thus, low-lying excited states observed in \nuc{40}{Si} probe 
predominantly neutron configurations, and 
the conclusions are not complicated by the proton degrees of freedom. 
Although a QRPA analysis of $\beta$-decay half-lives suggests \nuc{39-42}{Si} 
are deformed, confirmation by more direct methods is needed \cite{grevy04}.
In the regime of exotic nuclei, excitation energies are often the first 
observables accessible to experiments. Two complementary reactions---proton 
inelastic scattering and nucleon removal---have been used in the 
present study to probe the level scheme of \nuc{40}{Si}. 

The experiment was carried out at the National Superconducting 
Cyclotron Laboratory (NSCL) at Michigan State University.  
A primary beam of \nuc{48}{Ca} at 140 MeV/nucleon from the 
Coupled Cyclotron Facility (CCF) impinged upon a 
987 mg/cm$^2$ \nuc{9}{Be} target placed at the mid-acceptance position of the 
A1900 fragment separator \cite{a1900}.  Projectile fragments were then 
separated by the B$\rho$-$\Delta$E-B$\rho$ method to deliver a 
cocktail beam composed of several nuclear species, including \nuc{40}{Si}, 
to the target position of the S800 spectrograph \cite{Baz03}.  
The full acceptance of both the A1900 and the S800 analysis line, operated 
in focused mode, was used yielding a total momentum acceptance of 4\%.  
At the S800 target position, the RIKEN/Kyushu/Rikkyo liquid hydrogen (LH$_2$) 
target was placed into the path of the beam \cite{lh2}.  The spectrograph 
was set to accept projectile nuclei elastically scattered from protons in the 
target.  Due to the large momentum acceptance of the spectrograph, single- 
and multiple-nucleon removal channels were also observed for many 
of the incoming nuclear species.

\begin{figure}[tb]
\epsfxsize 8.5cm \epsfbox{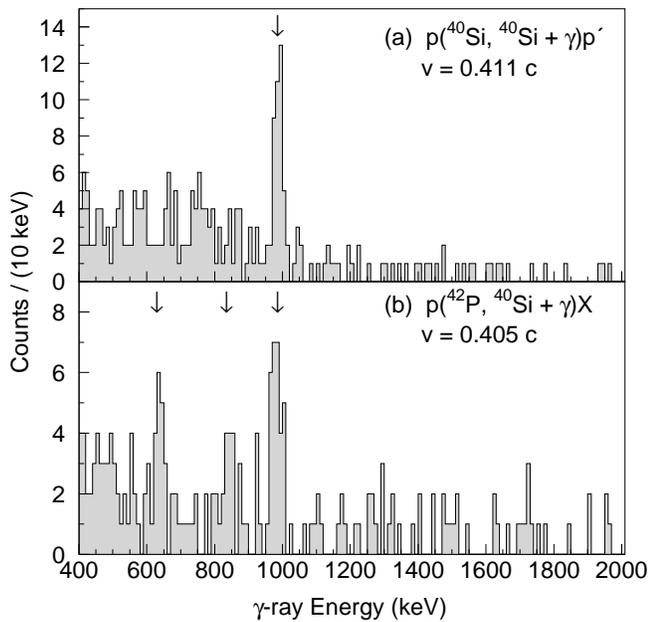}
\caption{ \label{gam}
Doppler-corrected $\gamma$-ray spectra observed in coincidence with 
the reactions $p(\rm{^{40}Si},\rm{^{40}Si}+\gamma)p^{\prime}$ and 
$p(\rm{^{42}P},\rm{^{40}Si}+\gamma)X$.}
\end{figure}

Prompt $\gamma$-ray decays of nuclei excited by inelastic scattering or 
left in an excited state by nucleon removal were detected by SeGA
(\underline{Se}gmented \underline{G}ermanium \underline{A}rray), 
an array of 32-fold segmented, high-purity Ge detectors \cite{Will01}.  
The array was configured around the LH$_2$ target in two rings with seven 
detectors at 37$^\circ$ and nine detectors at 90$^\circ$ relative to the 
beam axis.  Event-by-event Doppler reconstruction takes advantage of the 
detector segmentation, producing $\gamma$-ray spectra in the projectile 
frame $(v \sim 0.4c)$ with $\sim$3\% (FWHM) resolution at 1 MeV.

In thick-target, inverse-kinematics inelastic proton scattering, 
decay $\gamma$-rays are used to tag inelastic scattering to specific 
excited states \cite{iwa00}.  However, single- and multiple-nucleon 
removal reactions occur with comparable, or larger, cross-sections.  
Thus, identification of both the incoming projectile and the outgoing 
nucleus is required.  
The charge of each incident projectile was determined 
for each event using a Si-PIN detector placed upstream at the object position 
of the S800, and an ionization chamber in the focal plane of the S800 
determined the charge of each particle after the target \cite{Yur99}.  
The magnetic rigidity of each incident projectile was 
determined by measuring the 
dispersive angle at the intermediate image of the S800 analysis beam 
line using a pair of high-rate Parallel Plate Avalanche Counters.  
The magnetic rigidity of each particle exiting the target was 
determined by measuring 
the dispersive position of each particle in the S800 focal plane with a 
Cathode Readout Drift Chamber.  Time-of-flight for each 
particle was measured between timing scintillators at the exit of the 
A1900 and the focal plane of the S800.  Together, these kinematic 
measurements positively determined incoming mass and change in mass 
due to reactions.  

Excited states of \nuc{40}{Si} were populated by inelastic scattering of 
incoming \nuc{40}{Si} nuclei and by $pn$ removal from \nuc{42}{P}
upon collision with the LH$_2$ target.  Figure 1 shows projectile frame 
$\gamma$-ray spectra detected in coincidence with the reactions 
$p(^{40}\rm{Si},^{40}\rm{Si} + \gamma)p^{\prime}$ and 
$p(^{42}\rm{P},^{40}\rm{Si} + \gamma)X$.  The transition at 986(5) keV 
is the strongest $\gamma$-ray observed in each of the reaction channels and 
the only transition observed via inelastic scattering.  Its relative strength 
combined with the selectivity of $(p,p^{\prime})$ allow 
this transition to be assigned as the 
$2^+_1 \rightarrow 0^+_1$ transition.  Thus, the excitation energy of the 
first $2^+$ state in \nuc{40}{Si} is 986(5) keV.

\begin{figure}[tb]
\epsfxsize 8.5cm \epsfbox{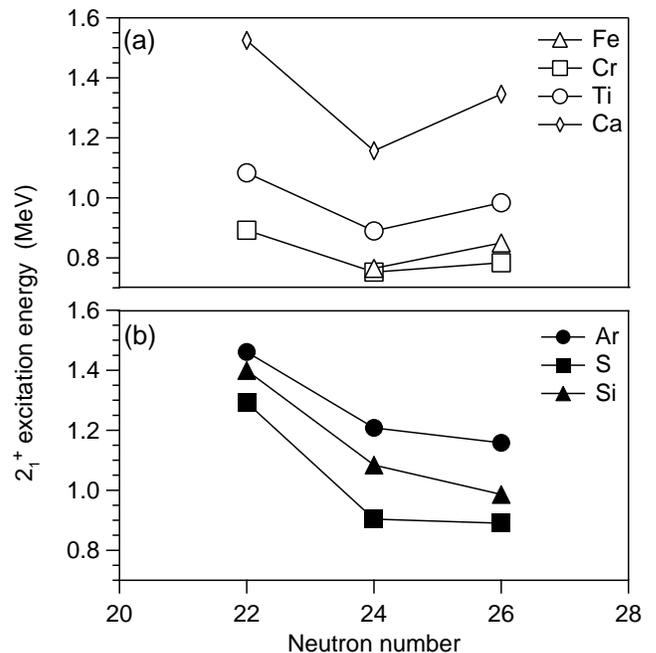}
\caption{ \label{trends}
Evolution of $2^+_1$ energy with neutron number for even-even nuclei 
with $Z\geq20$ (a) and $Z<20$ (b).}
\end{figure}

The $pn$ removal reaction leading to \nuc{40}{Si} also shows two weaker peaks 
at 638(5) and 845(6) keV, each with about half the intensity of the 
corresponding $2^+_1 \rightarrow 0^+_1$ transition.  The
energies of the 986(5) and 638(5) keV 
peaks agree with the observations of Grevy {\it et al.} \cite{grevy05}.  
The statistics obtained for the 638(5) and 845(6) keV transitions preclude
a discussion of whether these two $\gamma$-ray transitions
should be in parallel or form a cascade.  However, they are expected to 
decay from higher excited states, based on other even-even
nuclei that show signifcant $\gamma$-ray feeding of the $2^+_1$ when 
populated by the $pn$ removal reaction.  Thus, we propose
that at least one of these two lower-energy $\gamma$-rays is directly
feeding the $2^+_1$ state and that an excited state of
\nuc{40}{Si} lies at either 1624(7) or 1831(8) keV.

Figure 2 shows the energy of the first $2^+$ state, E($2^+_1$), versus 
neutron number ($N=22-26$) for even-even nuclei between silicon and 
chromium.  Data in this figure were taken from the ENSDF 
database \cite{ensdf06} and the current experiment.
Figure 2(a) illustrates the evolution of E($2^+_1$) 
in nuclei having large shell gaps for both $N=20$ and $N=28$.  
Changes in the excitation energy are dominated by the 
filling of a single-neutron orbital, $0f_{7/2}$; the trend is parabolic, 
reaching its minimum at midshell; and excitation energies are nearly 
symmetric about midshell, $N=24$.  In panel (b), this symmetry is lost.  
The lowering of first excited states in the neutron-rich sulfur and 
argon nuclei can be attributed, in part or in whole, to a narrowing of 
the $\pi(0d_{3/2}-1s_{1/2})$ gap \cite{reta97,cot98}.  The rise in 
collectivity in the sulfur and argon nuclei is, 
therefore, not sufficient to establish a breakdown of the $N=28$ shell 
closure.  The situation is quite different when considering the chain 
of silicon isotopes.  Because $Z=14$ is a strong shell closure, 
the decrease in $E_{2^+_1}$ between \nuc{38}{Si} and \nuc{40}{Si} can 
only be due to a reduction of the $N=28$ shell gap.

\begin{figure}[tb]
\epsfxsize 8.5cm \epsfbox{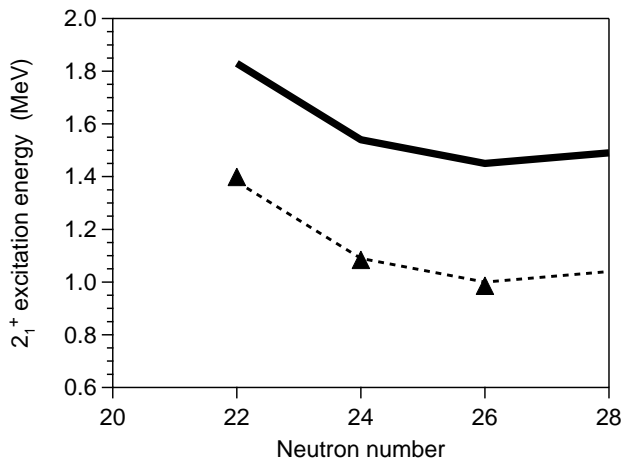}
\caption{ \label{smtrend}
Measured $2^+_1$ energies ($\triangle$) of \nuc{36,38,40}{Si} with 
corresponding shell-model predictions ({\bf ---}) described in the text.  
Shell-model predictions shifted lower by 450\ keV ({\bf- - -}) clearly 
reproduce the experimental values.}
\end{figure}

To examine the decline in $2^+_1$ energies across these neutron-rich 
silicon isotopes, large-scale shell-model calculations were performed 
with CMICHSM \cite{cmichsm} in a $\pi(sd)^{Z-8}~\nu(fp)^{N-20}$ model 
space using the updated interaction of Nowacki and collaborators 
\cite{reta97,num01}.  In Fig.~3, measured and predicted $2^+_1$ 
excitation energies are plotted versus neutron number.  While the 
evolution of excitation energies is in good agreement with experiment, 
the predicted values are consistently 450 keV higher than the measured 
values.  The first excited state of each nucleus with $15 \leq Z \leq 19$ 
and $N=22,24,26,28$ has been measured and compared with shell-model 
predictions using the same interaction \cite{reta97,num01,sohl02,gade06}.  
Agreement was quite good in all cases, and no other isotopic chain suffers 
from the large difference between prediction and theory observed in these 
silicon isotopes. Shell-model predictions reduced by an empirical offset 
of 450 keV are plotted in Fig.~3 as a dotted line.  These shifted 
predictions give an RMS error below 30 keV for \nuc{36,38,40}{Si} and 
predict the $2^+_1$ energy in \nuc{42}{Si} at 1050 keV.  
This would place the $2^+_1$ of \nuc{42}{Si} below that of \nuc{44}{S} 
in qualitative agreement with the observations of 
Grevy {\it et al.} \cite{grevy05}.  
Due to the pronounced proton sub-shell closure at $Z=14$, shell-model 
predictions indicate that the enhanced collectivity resulting from 
promotion of two neutrons across the $N=28$ shell gap will not result 
in strong deformation \cite{caur04}.

\begin{figure}[tb]
\epsfxsize 8.5cm \epsfbox{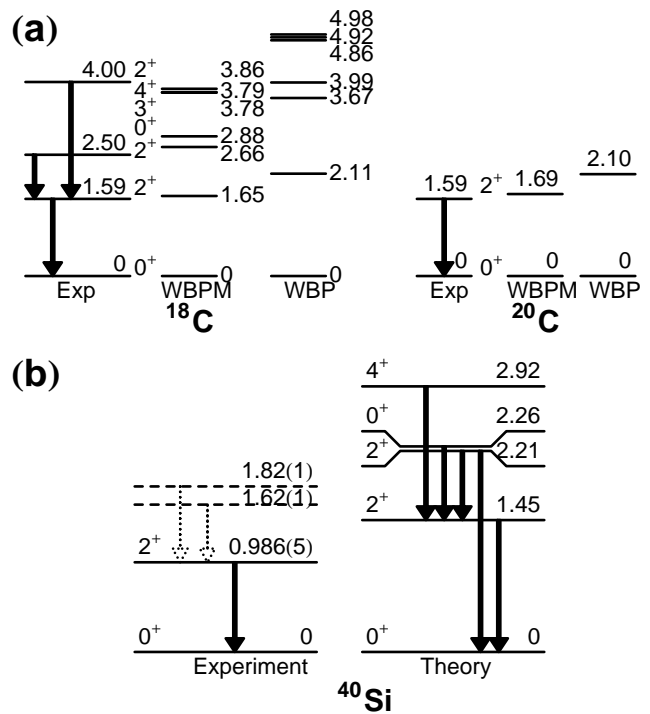}
\caption{ \label{lvlschm}
Panel (a) shows the experimental level schemes of $^{18,20}$C along with 
shell-model predictions using the WBP and WBPM interactions.  
With the WBPM, reducing the valence $n$-$n$ interaction strength 
leads to better agreement with experiment.  
Panel (b) shows measured and predicted levels in $^{40}$Si.  
The two $\gamma$ decay branches of the $2^+_2$ state 
are predicted to have comparable intensities.  
The existence at a low excitation energy of one or both of the states 
predicted above the first 2+ state would be consistent with the 
results of the present experiment.  
The experiment-theory shift is reminiscent of that seen in $^{18,20}$C.}
\end{figure}

A possible explanation of the disagreement between theoretical predictions 
and experiment in the silicon $2^+$ energies can be found in the 
neutron-rich carbon nuclei, which also have a 
strong proton sub-shell closure 
and initially showed a similar disagreement between theory and experiment.  
Measured and calculated spectra for \nuc{18,20}{C} are 
shown in Fig.~4(a) \cite{stan04,war92}.  By reducing the neutron-neutron 
($n$-$n$) interaction strength in carbon, the WBPM interaction reproduces 
the observed level spacing better than WBP.  In oxygen, the neutron orbital 
$1s_{1/2}$ lies 0.87 MeV above the $0d_{5/2}$ orbital at $N=9$. This gap 
grows with increasing neutron number due to the $n$-$n$ interaction 
resulting in a shell gap at $N=14$ which makes \nuc{22}{O} 
doubly magic \cite{bro05}.  In carbon, the neutron orbital $1s_{1/2}$ 
lies 0.74 MeV below the $0d_{5/2}$ orbital at $N=9$.  This orbital inversion 
leads WBP to predict $N=14$ is not a shell closure in carbon.  The weaker 
$n$-$n$ interaction used in WBPM causes the $\nu (1s_{1/2} - 0d_{5/2})$ gap 
to grow more slowly, thus reinforcing that no shell gap develops at $N=14$ 
in the carbon isotopes.  

If we compare the oxygen and carbon chains to the calcium and 
silicon chains, the similarities are striking.  Originally, shell-model 
calculations used the $1p_{3/2} - 0f_{7/2}$ neutron gap of \nuc{41}{Ca} with 
the $n$-$n$ interaction found in stable nuclei---which is known to widen the 
$N=28$ shell gap---to predict \nuc{42}{Si} as a doubly-magic nucleus 
\cite{reta97}.  Nummela {\it et al.} measured a reduction of this neutron 
gap in \nuc{35}{Si} leading to a new prediction of weakened magicity in 
\nuc{42}{Si}  \cite{num01}.  The current experiment shows that shell-model 
predictions give larger excitation spectrum spacing in silicon isotopes than 
is measured.  Reducing the $n$-$n$ interaction at $Z=14$ could correct the 
predicted energy spacing but would also imply that the $N=28$ gap grows more 
slowly in silicon than in calcium and that current shell-model predictions 
are, in fact, overestimating the size of the $N=28$ shell gap.  Experimental 
knowledge of higher lying states in the silicon isotopes is needed to fully 
examine this possible reduction in $n$-$n$ interaction strength.

Further support for a narrowing of the $N=28$ shell gap in \nuc{40}{Si} 
comes from the low energy inferred here for the second excited state, 
independent of any specific interpretation of the experimental 
level scheme.  In the shell model (compare Fig. 4(b)) the states 
predicted to decay through the $2^+_1$ level, via $\gamma$-rays of less 
than 1 MeV, arise from particle-hole excitations across the $N=28$ gap. 
Thus, their low excitation energy implies a reduced gap.  

To summarize, the $\gamma$ decays of excited states in \nuc{40}{Si} have been 
observed using the complementary techniques of inelastic scattering and 
$pn$ removal on a liquid hydrogen target.  An excitation energy of 
986(5) keV was measured for the $2^+_1$ state in \nuc{40}{Si}, and a 
second excited state at either 1624(7) or 1831(8) keV was deduced. The large 
proton sub-shell gap at $Z=14$ 
at $Z=14$ means that the evolution of excitation energies in the 
silicon isotopes is directly related 
to the narrowing of the $N=28$ shell gap.  The decline in $2^+_1$ energy from 
\nuc{38}{Si}, at midshell, to \nuc{40}{Si} can only be explained by a 
weakening of $N=28$.  The low energy of a second excited state in 
\nuc{40}{Si} is compatible with shell-model predictions of particle-hole 
excitations across the $N=28$ shell gap.  The $2^+_1$ energy trend 
from shell-model calculations taken with an empirical shift predicts the 
first $2^+$ energy in \nuc{42}{Si} will be below that of \nuc{44}{S}.

This work was supported by the National Science Foundation 
under Grants No. PHY-0110253, PHY-9875122, PHY-0244453, PHY-0555366, 
INT-0089581 and by the Japan Society for the Promotion of Science.

\end{document}